\documentclass[sn-mathphys]{sn-jnl}

\usepackage{graphicx}% Include figure files
\usepackage{dcolumn}% Align table columns on decimal point
\usepackage{bm}% bold math
\usepackage[mathlines]{lineno}% Enable numbering of text and display math
\usepackage[english]{babel}
\usepackage{setspace}
\usepackage{color}
\usepackage{float}
\usepackage[utf8]{inputenc}
\usepackage[T1]{fontenc}
\usepackage{etoolbox}

\space

\jyear{2022}%

\theoremstyle{thmstyleone}%

\theoremstyle{thmstyletwo}%

\theoremstyle{thmstylethree}%

\begin{document}

\title[]{Dispersion of Free-Falling Saliva Droplets by Two-Dimensional Vortical Flows}
% Force line breaks with \\

\author[1]{\fnm{Orr} \sur{Avni}}

\author*[1]{\fnm{Yuval} \sur{Dagan}}\email{yuvalda@technion.ac.il}

\affil[1]{\orgdiv{Faculty of Aerospace Engineering}, \orgname{Technion - Israel Institute of Technology}, \orgaddress{\city{Haifa}, \postcode{320003}, \country{Israel}}}

\abstract{
The dispersion of respiratory saliva droplets by indoor wake structures may enhance the transmission of various infectious diseases, as the wake spreads virus-laden droplets across the room. 
Thus, this study analyses the interaction between vortical wake structures and exhaled multi-component saliva droplets.

A self-propelling analytically-described dipolar vortex is chosen as a model wake flow, passing through a cloud of micron-sized evaporating saliva droplets. 
The droplets' spatial location, velocity, diameter, and temperature are traced, coupled to their local flow field.
For the first time, the wake structure decay is incorporated and analyzed, which is proved essential for accurately predicting the settling distances of the dispersed droplets.
The model also considers the non-volatile saliva components, adequately capturing the essence of droplet-aerosol transition and predicting the equilibrium diameter of the residual aerosols.
Our analytic model reveals non-intuitive interactions between wake flows, droplet relaxation time, gravity, and transport phenomena.

We reveal that given the right conditions, a virus-laden saliva droplet might translate to distances two orders of magnitude larger than the carrier-flow characteristic size.
Moreover, accounting for the non-volatile contents inside the droplet may lead to fundamentally different dispersion and settling behavior compared to non-evaporating particles or pure water droplets.
Ergo, we suggest that the implementation of more complex evaporation models might be critical in high-fidelity simulations aspiring to assess the spread of airborne respiratory droplets.
}


\maketitle

\section{Introduction}\label{sec:level1}
Direct contact with airborne exhaled respiratory droplets and aerosols is considered one of the main reasons for the rapid spread of viral diseases. 
During respiratory events of an infected individual such as talking, coughing, or sneezing, a variable amount of virus-laden droplets is released into the air.
Exposure to such contaminated droplets might infect nearby susceptible persons, transmitting the disease and increasing the risk of it spreading rapidly and uncontrollably~\citep{Organization2021}. 
Possible routes of transmission that have been widely accepted are the "airborne droplet" and "airborne aerosol" routes~\citep{Li2021basic,Chen2020}. The first involves the deposition of larger droplets on the mucosal surfaces (e.g., eyes, nose, mouth), while the second refers to the inhalation of finer droplets by the infected individual.
Thus, the persistence and dispersion of sub-micron-sized aerosols are of great concern, especially due to implications regarding public gatherings in closed environments~\citep{Tellier2019}.
However, the extent to which airborne mechanisms are responsible for the vast COVID-19 outbreak remains debated~\citep{chagla2020re,asadi2020coronavirus, smith2020aerosol}.

Past studies on the hydrodynamics of exhaled respiratory droplets show that for most scenarios where human coughing and sneezing into the quiescent environment are involved, the settling distances of largest-scale droplets do not exceed $1.5-2m$ ~\citep{Xie2007,Wang2020}.
Hence, the 2-meter rule was set as a standard for social distancing, recommended by most countries during the outbreak of COVID-19.
Nevertheless, different flow conditions may spread relatively large exhaled droplets farther than 2 meters \citep{Balachandar2020}.
In such cases, the effectiveness of social distancing, barriers, and ventilation intended for suppressing the spread of viral diseases might vary significantly, as the dispersion of droplets relies on the specific flow condition.
The conditions, in turn, depend on multiple factors such as ventilation rates, air conditions, moving objects, and flow barriers.
On the other hand, flow conditions generated by expiratory events may lead to the persistence of virus-laden droplets in the flow, thus producing regions in which the risk of infection via the "airborne droplet" route increases dramatically \citep{Ng2021}.
This interrelated dependency between the settling distance, virus exposure, and the flow conditions may lead to inconclusive and ineffective recommendations by health regulators \citep{Balachandar2020}. 
For example,~Wang et al.~\cite{Wang2021a} computed the density of virus copies moving past a defined control area, rather than accounting for the number of droplets.
This approach produced useful exposure maps and highlighted the risk in short-range exposure,  underestimated by the current guidelines.

Recent research efforts studying the hydrodynamic dispersion of exhaled respiratory droplets focus on either speaking, coughing, or sneezing into a quiescent environment~\citep{bourouiba2014violent,chong2021,Liu2021}, or assume a well-mixed homogeneous environment in which droplets and aerosols spread equally in space~\citep{Bazant2020,bazant2021guideline}. 
%Others examined specific cases such as the efficiency of different masks in blocking the spread of droplets~\citep{mittal2020mathematical, verma2020visualizing} and the dispersion of exhaled droplets during outdoor and indoor activities in different flow environments~\citep{zhang2021disease,liu2021simulation,nazari2021jet}.

Almost all indoor activities induce vortical flow structures. Ventilation systems, flow over barriers, or even indoor movements of small objects, will shed unsteady vortices that may affect the spread of droplets and aerosols.
The indoor airflow due to human movement and ventilation systems is rather complex \cite{Luo2018a}, and was found to affect the spread of airborne particles \cite{Wang2015}.
Two counter-rotating vortical structures may constitute a significant part of these wake flows, depending on the geometric constraints and environmental conditions.
Comparable vortex dipole structures were observed in several geophysical flows \citep{Ahlnas1987,Haines1987}, in the wake of aircraft \citep{Jugier2020}, and confined plasmas \citep{Nycander1990}.
Once created, these flow structures self-propel through space and sustain for relatively long periods, causing them to influence various transport phenomena. 

Renzi and Clarke~\cite{renzi2020} have investigated the role of vortex rings in the settling of exhaled respiratory droplets.
Using a theoretical approach and assuming Hill's equation as the carrier flow field, they found that vortex flows may delay the settling time of particle suspended within an exhaled jet and thus enhance its displacement.
Moreover, Chong et al.~\cite{chong2021} and Ng et al.~\cite{Ng2021} investigated the role of humid, turbulent puff ejected during a typical respiratory event on the extension of respiratory droplets' lifetime.
The studies reported that small droplets may be captured inside vortical structures constituting turbulent puffs.
This entrapment leads to a significant increase in the droplets' lifetimes, implying that smaller droplets might be transported much further than expected during respiratory events. 
Liu et al.~\citep{Liu2021a} explored the fluid dynamics of a turbulent, ejected laden puff resulting from a cough or sneezing. Large-eddy simulations were used to analyze the carrier fluid phase, while a point-particle Euler–Lagrange approach was chosen for tracking respiratory droplets. Notably, they observed the detachment of a small vortex ring-like structure that advanced to relatively large distances while carrying along part of the suspended droplets.
%Moreover, \citeauthor{Cummins2020}\cite{Cummins2020} have studied the dynamics of spherical droplets in the presence of a source-sink pair potential flow field. A mathematical analysis was presented, offering a theoretical model for droplet dispersion using an analytically described carrier flow field.
%The study demonstrated the interactions between particles, gravity, and carrier potential flow. The study has revealed an intermediate range of droplet diameters for which their horizontal distances are minimized.

The stability of laminar and turbulent flows may be affected by large recirculation regions ~\citep{dagan2016, taamallah2019helical, chakroun2019flamelet, DAGAN2019368}. Our recent studies reveal the complex interactions of such unstable flows and evaporating reacting sprays~\citep{dagan2015dynamics}.
Mathematical descriptions of such interactions were derived to solve the coupled vortex-particle system~\citep{daganILASS2016, daganILASS2017, daganSimilarityFlames2018}, and may also assist in understanding and interpreting results from numerical simulations~\citep{dagan2017particle}. 

In our recent study \citep{Dagan2021}, we used the two-dimensional Lamb-Chaplygin dipole solution to examine the effects of vortex dipoles on the dispersion and settling distance of non-evaporating particles. 
However, we did not consider the heat and mass transfer between the droplet and the ambient humid air.
The ensuing research \citep{Avni2021} has expanded this configuration, serving as a generalized model for respiratory droplets suspended in the ambient air, and accounted for the mass transfer. 
The effect of evaporation was analyzed and studied by comparing the trajectories and settling distances of the evaporating droplet and a solid particle.
 The existence of optimal conditions for maximum displacement was suggested, where the droplet translation reaches up to an order of magnitude larger than the vortex core length scale. 

The influence of droplet evaporation rates on the airborne droplet lifetime and settling distances have been studied extensively for close to a century \citep{Wells1934,Xie2007}.
Subsequently, the effect of air and droplet properties such as ambient temperature, diameter, local velocity, and relative humidity has been established as critical for the accurate analysis of the persistence and dispersion of liquid droplets \citep{DeOliveira2021}.
In addition to the air and droplet properties, the droplet's composition should be considered when aiming to predict its persistence.
This realization consolidates when analyzing the lifetime of a saliva droplet, which is particularly important for the assessment of the COVID-19 spread and transmission \citep{Xu2020}.
Nonetheless, most studies on the hydrodynamics of exhaled respiratory droplets modeled them as water droplets.
Although human saliva's main component is water ($\sim 99\%$), it is a complex biological fluid comprised of various proteins, organic compounds, and inorganic salts \citep{Sarkar2019a}.
A model saliva formulation had been developed by Sarkar et al.~\citep{Sarkar2019a}, offering a good estimation of the bulk human saliva properties from electrostatics and viscosity perspectives.
The formulation includes seven different salts, urea solutions, and varying concentrations of mucins. 

As the droplet's water content evaporates, it may transition from a saliva droplet to a solid residue, featuring a unique behavior; regardless of its initial diameter and the ambient conditions, the final diameter is set to be roughly 20\% of the initial droplet diameter\citep{Stiti2021,Basu2020}.
This aerosol might remain stable for hours and would nearly suspend in the ambient air due to its small mass\citep{Lieber2021}.
Since the aerosol deposition and lifetime is much longer when compared to the droplet from which it originated, viscous effects may come into play and alter the dynamic interaction between the particle and the flow carrying it.

Hence, the objective of the present study is to extend and solidify our recent works~\citep{Dagan2021,Avni2021}, and investigate the effects of viscid, vortical wake structures on the dynamics and deposition of evaporating saliva droplets, free-falling in indoor environments. 

A mathematical formulation for the wake flow is presented in Sec.\ref{sec:flow}. In Sec.\ref{sec:level4}, we describe the Lagrangian model, spatially tracking the droplet's velocity, mass, and temperature.
Verification of our Lagrangian model, as well as a validation of the numerical procedure, is introduced in Sec.\ref{sec:v&s}.
The results are presented in Sec.\ref{sec:results} for a cloud of exhaled droplets in the vicinity of a vortical flow structure, where the settling distances and trajectories of the saliva droplets are analyzed and studied.
Finally, Sec.\ref{sec:conc} includes concluding remarks and a brief outlook.

% SECTION 2
\section{\label{sec:math} Mathematical model}

We consider a dilute cloud of discrete micron-sized droplets free-falling in initially still, undisturbed air.
A wake flow structure, generated by a movement of a nearby object, encounters the cloud and consequently might carry droplets downstream. Since the droplet cloud is dilute, and small relative to the wake structure, we assume the unsteady motion of the droplets does not affect the flow field. Furthermore, we shall discount any potential interactions between the droplets, including collision and coalescence, and assume that each droplet does not affect the properties governing the dynamics of surrounding droplets.
The extent to which the wake flow disperses the cloud will be attained by utilizing a Lagrangian approach, where the single droplet's spatial location $\vec{x}_p$, velocity $\vec{u}_p$, diameter $d_p$, and temperature $T_p$ will be traced and coupled to the local wake flow field $\vec{u}_{f}$. 
The equations for the carrier flow and the Lagrangian droplet are presented as follows.

\subsection{\label{sec:flow} Wake flow}

\subsubsection{\label{sec:ideal} Lamb-Chaplygin dipole}

 In favor of gaining physical insight into the complex interactions between a droplet and the flow field, we approximate the flow field with a two-dimensional analytical-described Lamb-Chaplygin dipole, serving as a preliminary model for realistic wake flows. The carrier flow is fully laminar and thus does not incorporate the dispersion caused by fine-scale turbulence. 
The self-propelling dipolar vortex is of a constant radius $a$ and moves at a velocity $U$ in quiescent air.
A solution for this idealized flow was derived by Lamb \citep{Swaters1988} and Chaplygin~\citep{Chaplygin2007} for an inviscid fluid, postulating a linear relation between the vorticity field and the stream function $\omega=k^2\psi$.
For radial coordinates $(r,\theta)$ positioned at the center of the vortex core in a frame of reference moving with the vortex, the velocity field inside the core is
\begin{equation}
    \begin{split}
        &u_{r,0}=-2a U_0 \cos(\theta) \frac{J_1(\frac{br}{a})}{brJ'_1(b)}\,,\\ &u_{\theta,0}=-2U_0 \sin(\theta) \frac{J'_1(\frac{br}{a})}{J'_1(b)},\;\;\;\;r\leq a,    
    \end{split}
    \label{eq:flowvel0}
\end{equation}
whereas the outer flow field solution is a potential flow past a cylinder.
$J_1, J'_1$ and $b$ denote the first-order Bessel function of the first kind, its derivative, and its smallest non-trivial root $b=ka=3.8317$, respectively.
The vorticity field varies inside the core, as the location of the maximal vorticity points are $(x=0,y=\pm 0.48a)$, and their values in terms of the translation velocity and core radius are 
\begin{equation}
    \omega_m=11.09Ua^{-1}. 
    \label{eq:dec}
\end{equation}
A streamline representation of the Lamb-Chaplygin vortex is illustrated in Fig. \ref{fig:strm}(a), for a normalized $(\bar{x},\bar{y})$ moving frame of reference, alongside the vorticity field intensity distribution.

\subsubsection{\label{sec:decay} Viscosity Effects}
Swaters \citep{Swaters1988} described the viscid, adiabatic decay of a Lamb-Chaplygin dipole using a perturbation theory. The analysis unveiled that the inner vorticity field undergoes an exponential decay;
\begin{equation}
    \omega_m(t)=\omega_{m,0}\exp\left(\frac{-t}{\tau_d}\right), 
\end{equation}
where $\omega_{m,0}$ is the initial maximal vorticity of the ideal flow solution, whereas the viscid decay time is   
\begin{equation}
    \tau_d=\frac{a^2}{b^{2}\nu_f}\approx 0.07\tau_{d,v}. 
\end{equation}

The modulation suggested by Swaters viscid decay was studied and validated experimentally by Flór et al.~\citep{Flor1995}; they analyzed the vortex decay of a thin-layered, quasi-two-dimensional dipole in both vertical and radial directions.   
The study confirmed the validity of the predicted decay due to the spread of the viscid layer experimentally and found its timescale to be $t\approx\tau_{d,v}m$.
Therefore, in the limit $t<\tau_{d,v}$ we may account for the influence of the viscid decay by multiplying the inviscid flow-field solution by a decay correction factor,
\begin{equation}
    \vec{u}_{f}(x,y,t)=\vec{u}_{f,0}(x,y)\exp\left(-t/\tau_d\right).
    \label{eq:ff}
\end{equation}
Thus, we use Eq. \ref{eq:ff} as an approximation of an indoor wake flow, dispersing a cloud of free-falling droplets exhaled in its vicinity.

\begin{figure*}
    \centering
  \includegraphics[width=0.95\linewidth]{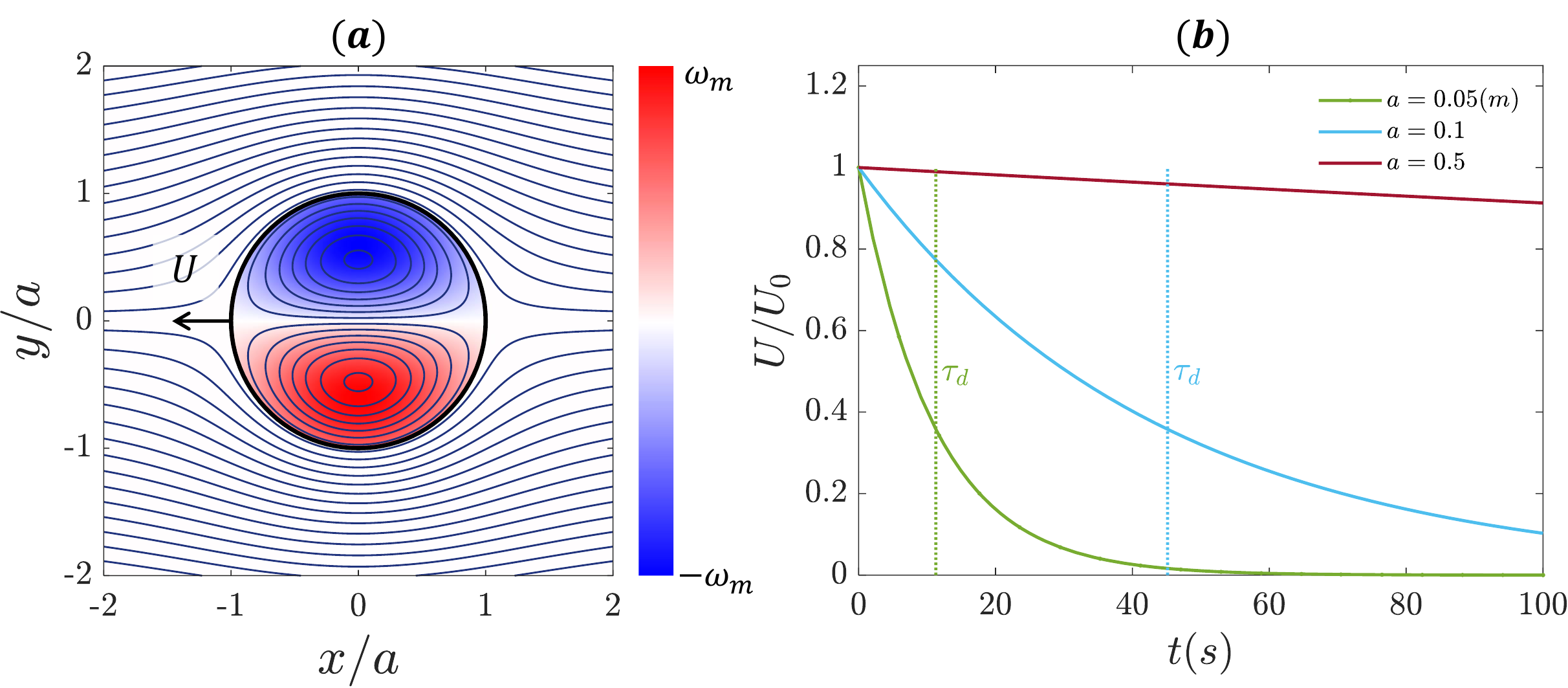}
  \caption{(a) Lamb-Chaplygin vortex dipole streamlines and vorticity field $\omega(\bar{x},\bar{y})$, illustrated at the normalized frame of reference fixed to the dipole center. (b) Comparison of the dipole's rate of viscid decay for different core sizes, represented by the ratio between the translation velocity and its initial value. vertical dotted lines demonstrate the characteristic decay times for each core size.}
  \label{fig:strm}
\end{figure*}

Fig. \ref{fig:strm}(b) demonstrates the characteristic decay of the velocity field for different core radii. One may notice the relation between the core size and the rate of decay; while a $a=0.05(m)$ vortex dissipates almost entirely after 50 seconds, a  $a=0.5(m)$ vortex would lose less than 10\% of its initial velocity. Therefore, when studying droplet-vortex interactions, we can compare the droplet relaxation time with the vortex lifetime and may exclude small vortices from the analysis, if their lifetimes are much shorter than droplet relaxation times, and thus will not influence the droplet's trajectory. 

\subsection{\label{sec:level4} Single droplet Lagrangian equations}

\subsubsection{\label{sec:momentum} Momentum equation}
The general form of the equations of motion for small particles in nonuniform, unsteady flows was derived by \cite{Maxey1983}, taking into consideration gravity, drag, virtual mass, and the Basset "history" force.
This study concerns a small water droplet motion in the air, where the particle-medium density ratio is large, and the particle characteristic length is much smaller compared to the flow integral length. 
 Hence, we may postulate that the magnitude of the forces due to undisturbed flow, virtual mass, Faxen's drag correction, and particle history terms are of a couple of orders smaller than the magnitude of the drag and gravity forces.
Additionally, although the carrier flow field is rotational, assuming the droplet is small may allow us to neglect rotational inertia's influence over the droplet's dynamics. Now, we may reduce the general form of the equations to the following:
\begin{equation}
    \frac{d\vec{x_p}}{dt}=\vec{u_p},
    \label{eq:location}
\end{equation}
\begin{equation}
    \frac{d\vec{u_p}}{dt}=18f\frac{\nu_{f}\rho_{f}}{\rho_{p}d_p^2}
    (\vec{u_f}-\vec{u_p})+\vec{g}-\frac{d\vec{u_f}}{dt},
    \label{eq:moment}
\end{equation}
where $\rho_{f}$ is the fluid (air) density, $\rho_{p}$ is the droplet density, $f$ is the drag factor, $\vec{g}$ is the gravitational acceleration, and $\vec{x_p}$, $\vec{u_p}$ are the droplet location and velocity vectors in the vortex frame of reference.
Since the chosen frame of reference is non-inertial, a fictitious force $d\vec{u_f}/dt$ is introduced to Eq.\ref{eq:moment}, accounting for the deceleration of the vortex translation velocity due to viscid dissipation.
The drag factor $f$, defined as the ratio of the drag coefficient to Stokes drag coefficient, was correlated by several authors as a function of the droplet's relative Reynolds number $Re_p=d\lvert\vec{u_f}-\vec{u_p}\rvert/\nu_f$.
For $Re_p$ up to 800, covering the physically-reasonable velocities of indoor airborne particles, the drag factor was found by Schiller and Naumann \citep{Crowe2011} to be:
\begin{equation}
    f=1+0.15Re_{p}^{0.687},
\end{equation}

\subsubsection{\label{sec:mass_eq} Mass equation}
The diameter and temperature of the Lagrangian droplet are governed by mass and heat and transfer processes, correspondingly. For common indoor environments, it is reasonable to assume that the exhaled respiratory saliva droplet will either lose mass and evaporate, or stay at equilibrium with its environment. 

Kulmala \citep{Kulmala1989,Kulmala1991a} have formulated the mass transfer at the droplet-air interface for a quasi-stationary case, assuming a zeroth-order mass fraction profiles around the droplet, the air is an ideal gas, and the droplet-air interface to be saturated.
Although diffusivity was considered the sole transfer mechanism, an expansion of the solution for ventilated droplets can be implemented using the Sherwood dimensionless number $Sh$, defined as the ratio of total mass transfer to the purely diffusive flux.
In terms of droplet diameter, the mass equation is
\begin{equation}
    \frac{d(d_p)}{dt}=4ShC_T\frac{M_{w,v}D_{\infty}p_{\infty}} {RT_{\infty}\rho_p d_p} \ln\left(\frac{1-\bar{p}_{v,d}}{1-\bar{p}_{v,\infty}}\right),
    \label{eq:mass}
\end{equation}
where $M_{w,v}$ is the molecular weight of the vapor, $R$ is the universal gas constant, $D_{\infty}$ is the ambient binary diffusion coefficient, while $p_{\infty}$, $\bar{p}_{v,\infty}=p_{v,\infty}/p_{\infty}$, and $\bar{p}_{v,p}=p_{v,p}/p_{\infty}$ are the air ambient pressure, ambient partial vapor pressure, and partial vapor pressure at the air-droplet interface, correspondingly.

The term $C_T$ in Eq. \ref{eq:mass} is the diffusion coefficient temperature dependence factor, which is given by
\begin{equation}
    C_T=\frac{T_{\infty}-T_{p}}{T_{\infty}^{\mu-1}} \space \frac{2-\mu}{T_{\infty}^{2-\mu}-T_{p}^{2-\mu}}.
    \label{eq:Ct}
\end{equation}
The term $\mu$ in Eq. \ref{eq:Ct}~is a substance-specific constant, which ranges between $1.6<\mu<2$ for most substances. 
For water droplets in room conditions, the temperature depression $T_{\infty}-T_{p}$ is relatively small, and thus the deviation in the diffusion coefficient predicted by the factor $C_T$ is smaller than 3\% \cite{Kulmala1989}. Hence, we may set $C_T$ to equal 1, and neglect the variation in the diffusion coefficient.

Eq. \ref{eq:mass} reveals that the mass diffusion from the droplet is governed by the ratio between the vapor's partial pressure at the evaporation interface and the partial pressure at the ambient, and thus an accurate estimation of both is in need. 
We take the ambient vapor pressure $p_{v,\infty}$ as a function of the ambient temperature and the relative humidity of the air
\begin{equation}
    p_{v,\infty}=\phi p_{sat}(T_{\infty}),
    \label{eq:vapor_infty}
\end{equation}
and calculate the temperature-dependent saturation pressure $p_{sat}(T)$ using the well-known Goff-Gratch formula \citep{goff1957saturation}.
For a droplet of pure water, we may assume the vapor pressure at the droplet surface $p_{v,p}$ is the vapor pressure corresponding to the droplet temperature $p_{v,p}=p_{sat}(T_p)$, as we considered the air film around the droplet to be fully-saturated.

Accurate modeling of the evaporation of saliva can be rather complex, and thus most studies simplify the saliva as an ideal solution of dissolved inorganic salts~\citep{Xie2007,Liu2017a,Stiti2021}.
This simplification allows for the estimation of the vapor pressure using Raoult's law,
\begin{equation}
    p_{v,p}=\chi_w p_{v,sat}(T_p),
    \label{eq:raoult}
\end{equation}
as $\chi_w$ is the water mole fraction in the droplet. Assuming $N$ different precipitates are dissolved in the solution and that only water molecules escape the droplet during the evaporation, the water mole fraction is given as 
\begin{equation}
    \chi_w=\frac{n_w}{n_w+n_s}=\left [ 1 + \frac{M_{w,H_2O}}{\rho_w} \left ( \sum_{j=1}^N \frac{C_{j,0}i_j}{M_{w,j}} \right ) \left (\frac{d_{p,0}}{d_p} \right )^3 \right ] ^{-1},
    \label{eq:mole_fr}
\end{equation}
where $C_{j,0}$ is the precipitate's initial molar concentration, $i_j$ is the precipitate ion factor (the number of dissolved ions due to the disassociation of one molecule of precipitate), $M_{w,H_2O}$ is the water molar weight, $M_{w,j}$ is the precipitate molar weight and $d_{p,0}$ is the droplet's initial diameter.

Seeing that evaporation is limited by the vapor pressure ratios, the decrease of the water mole fraction inside the droplet as it shrinks may result in earlier supersaturation of the air film around the droplet, thus retarding the evaporation process.
Evaporation might also lead to salt crystallizing inside the droplet, as well as permeable crust forming around the droplet \citep{Basu2020}. However, we do not consider the influence of these phenomena, as they are found to be relatively insignificant for saliva evaporation under typical indoor conditions \citep{Lieber2021,Basu2020}, and treat the saliva as an ideal mixture of inorganic salts dissolved in water.
 Substituting Eq. \ref{eq:vapor_infty}-\ref{eq:mole_fr} into Eq. \ref{eq:mass} and approximating the vapor pressure ratios using the Clausius-Clapeyron relation, we reveal that the equilibrium droplet size is dependent on the ambient conditions, the droplet temperature, and the relative humidity;
\begin{equation}
    \frac{d_{p,eq}}{d_{p,0}}=\left \{ {\frac{2 \rho_w C_{NaCl} } {M_{w,H_2O} M_{w,NaCl}} \left ( {\phi\exp\left[{\frac{M_{w,H_2O}h_{fg}}{R}\left(T_{\infty}^{-1}-T_{p}^{-1}\right)}\right]-1} \right)^{-1}} \right \}^{1/3}.
    \label{eq:eq_dia}
\end{equation}

Effros et al.~\citep{Effros2002} estimated the concentrations of ions in human saliva as approximately $300 mM$ of both anions and cations.
We treat the saliva as a saline solution with  $9 g/l$ NaCl mass concentration, equivalent to the concentration of various ions found by Effros et al. in sampled saliva.
Under this assumption Eq.\ref{eq:eq_dia} predicts the equilibrium diameter is roughly 20\% of the initial diameter for indoor conditions, which correlates well with the available experimental results \citep{Lieber2021,Basu2020,Stiti2021}. 

\subsubsection{\label{sec:mass} Energy equation}
Both sensible heat stored in the droplet, heat advection due to mass transport, conduction, and convection at the droplet surface are the mechanisms controlling the droplet temperature. We derive the energy conservation equation for a single droplet assuming a uniform temperature distribution as
\begin{equation}
    \frac{dT_p}{dt}=\frac{3h_{fg}}{c_{p}d_p} \left(\frac{d(d_p)}{dt}- \frac{  4Nuk_f}{h_{fg}\rho_p}\frac{T_p-T_{\infty}}{d_p}\right),
    \label{eq:heat}
\end{equation}
where $h_{fg}$ is the heat of vaporization, $c_{p}$ is the droplet heat capacity, $k_f$ is the fluid heat conductivity, and $Nu$ is the Nusselt dimensionless number.

By applying a dimensional analysis based on the similarity of the mass and heat equation, we may determine $Sh=Sh(Re_p,Sc),Nu=Nu(Re_p,Pr)$, where $Sc=\nu_f/\rho_f D_{\infty}$ and $Pr=c_{p,f}\nu_f/\rho_f k_f$ are the fluid Schmidt and Prandtl numbers, respectively.
Correlations for both Sherwood and Nusselt numbers were found experimentally for moderate Reynolds \citep{Fuchs1959},
\begin{equation}
    \begin{split}
        &Sh=1+0.276Re_p^{1/2}Sc^{1/3}\,;\\ &Nu=1+0.276Re_p^{1/2}Pr^{1/3},
    \end{split}
\end{equation}
and will be adopted in the present study.

\section{Setup and Validation} \label{sec:v&s} 

\subsection{Verification and Validation} \label{sec:conf}

Unlike the carrier flow, which is described analytically by Eq.\ref{eq:ff}, the coupled ordinary differential equation system is solved numerically.
We track the droplet's spatial location by solving Eq.\ref{eq:location} and find its Lagrangian velocity by solving Eq.\ref{eq:moment}. This equation, in turn, is coupled to Eq.\ref{eq:mass} and Eq.\ref{eq:heat}, describing droplet diameter and temperature, correspondingly. The coupled system is solved simultaneously using an RK4 method with a variable time step.

The accuracy of our numerical scheme was verified with the results of Xie et al.~\cite{Xie2007}, who had developed a semi-analytic model and investigated the evaporation and movement of droplets expelled during respiratory activities.
Moreover, we validate the model using the experimental findings of Stiti et al. \citep{Stiti2021} and Basu et al.~\citep{Basu2020}, who investigated, independently, the evaporation of levitating stationary saliva droplets under various environmental conditions, detailed in table~\ref{tab:exp_conditions}.

\begin{figure}
    \centerline{\includegraphics[width=0.6\linewidth]{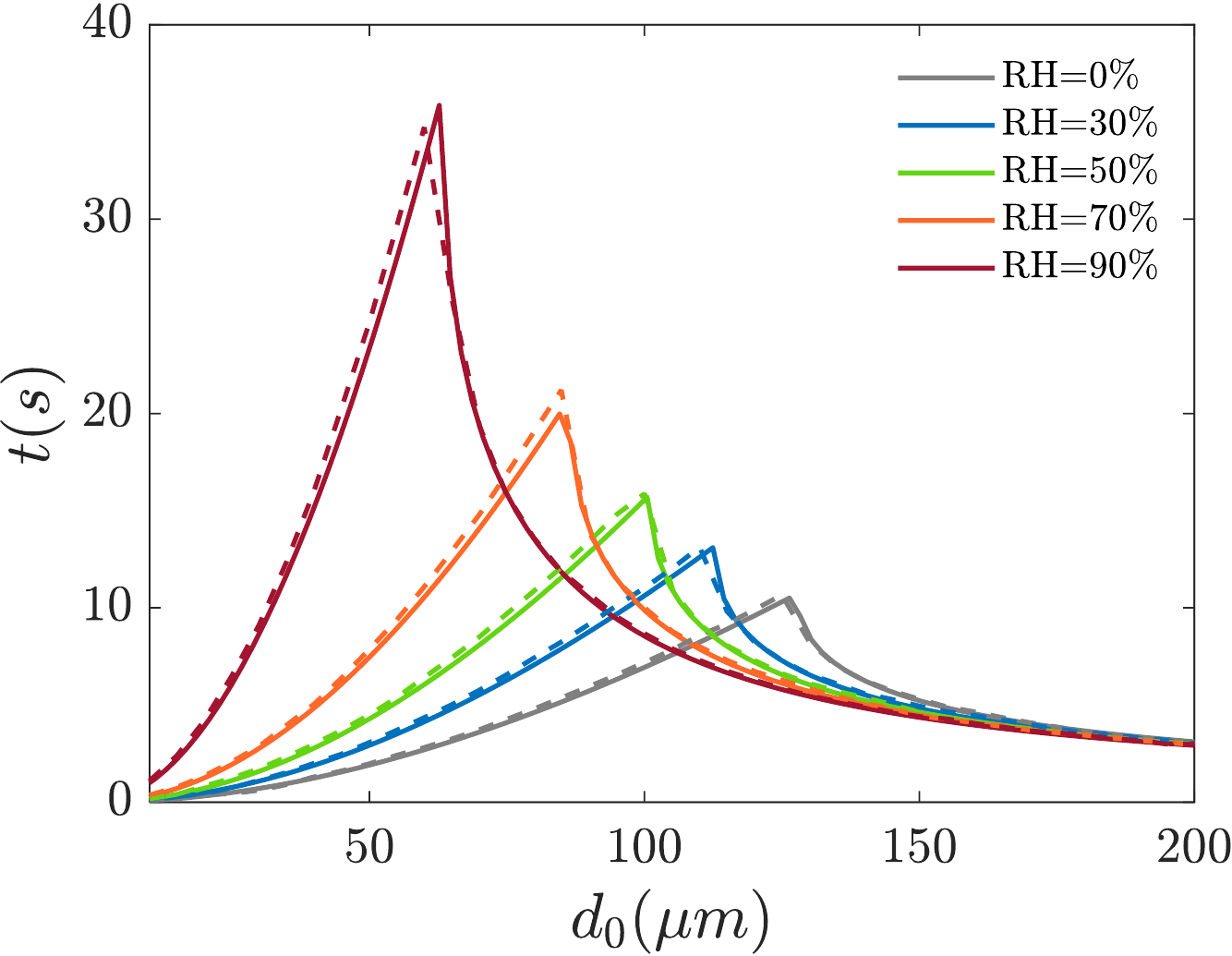}}
    \caption{Airborne lifetime of a $T_{d,0}=306K$ water droplet free-falling in quiescent air ($T_{\infty}=291K$) and various relative humidities, as a function of its initial diameter. Solid lines represent the current study predicaments, and dashed lines are modeling results of Xie et al.~\cite{Xie2007}.}
    \label{fig:wells}
\end{figure}

Some of the main findings of Xie et al.'s study are a revision of Wells's evaporation–falling curves for droplets with varying initial diameters, free-falling from 2 m to the ground in quiescent air of ambient temperature $T_{\infty}=291K$ and different relative humidities.
Comparison between the published results and the solution offered by the current model are plotted in Fig. \ref{fig:wells}.
A good agreement between the current study and Xie et al.'s numerical results was achieved. The influence of the ambient air properties on the heat and mass transport out of the droplet is clearly shown here. As expected, the evaporation of droplets is significantly slower in moist air than in dry air, and thus they accumulate more inertia and fall quicker. On the other hand, under such conditions, the airborne lifetime might be enhanced due to slower evaporation, delaying the drying of the droplet.

\begin{figure}
    \centerline{\includegraphics[width=0.7\linewidth]{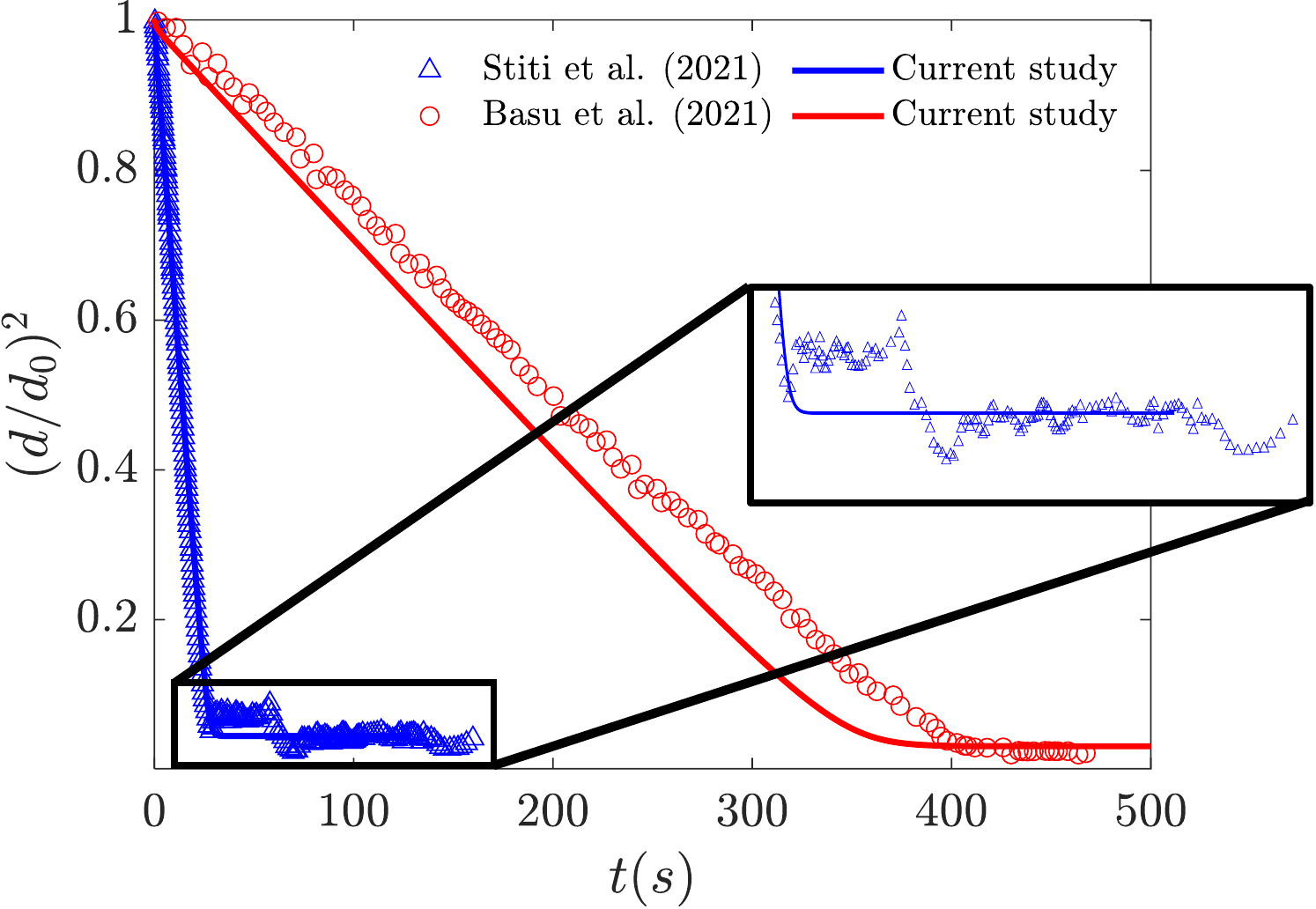}}
    \caption{Comparison of evaporation rates of levitating saliva droplet measured by Stiti et al.~\cite{Stiti2021} (blue triangles), Basu et al.~\citep{Basu2020} (red circles), and the numerical solution of the current study (solid lines).}
    \label{fig:val}
\end{figure}

Fig. \ref{fig:val} presents comparisons between the experimental data and the proposed numerical solution. The results fit the experiments reasonably, thus highlighting the validity of the numerical procedure.
Initially, the surface area of the droplet decreases linearly and follows the classical $d^2$ law.
However, the evaporation process quickly diverges from the linear trend when the normalized surface area $(d/d_0)^2$ reaches a value of roughly 0.05.
A sudden stop of the drying process is observed, as both the measured and calculated droplet's size remains constant.
The stagnation is illustrated in the inset of Fig. \ref{fig:val}.
As expected, the presence of NaCl ions in the modeled saliva solution lowers the vapor pressure outside the droplet, and thus prevents its complete evaporation.
The calculated residual diameters are in the range of $d_{p,r}/d_0=0.23-0.24$, and are found to be independent of the initial droplet size and vary slightly with the ambient conditions, in agreement with previous studies \citep{Basu2020,Lieber2021}.
Although we chose a rather simplistic approach it is evident that our model, without any pre-calibration to specific conditions, captures the essence of the droplet-aerosol transition and predicts the aerosol equilibrium diameter correctly.
Therefore, we may utilize it to approximate the evaporation of a saliva droplet, although it does not account for the complex composition of human saliva. 

\begin{table}
\centering
    \begin{tabular}{c@{\hspace{0.5in}}c@{\hspace{0.2in}}c@{\hspace{0.2in}}c}
    Study & $d_0~(\mu m)$ & $T_{\infty}~(^{\circ}C)$ & $RH~(\%)$  \\ \hline
   Stiti et al.~\citep{Stiti2021} & 121 & 22 & 55 \\ 
    Basu et al.~\citep{Basu2020} & 550 & 28 & 41 \\ 
    \end{tabular}
    \caption{Levitating saliva experimental conditions.}
    \label{tab:exp_conditions}
\end{table}

\subsection{Model Configuration} \label{sec:mod}

\begin{figure}
    \centerline{\includegraphics[width=0.7\linewidth]{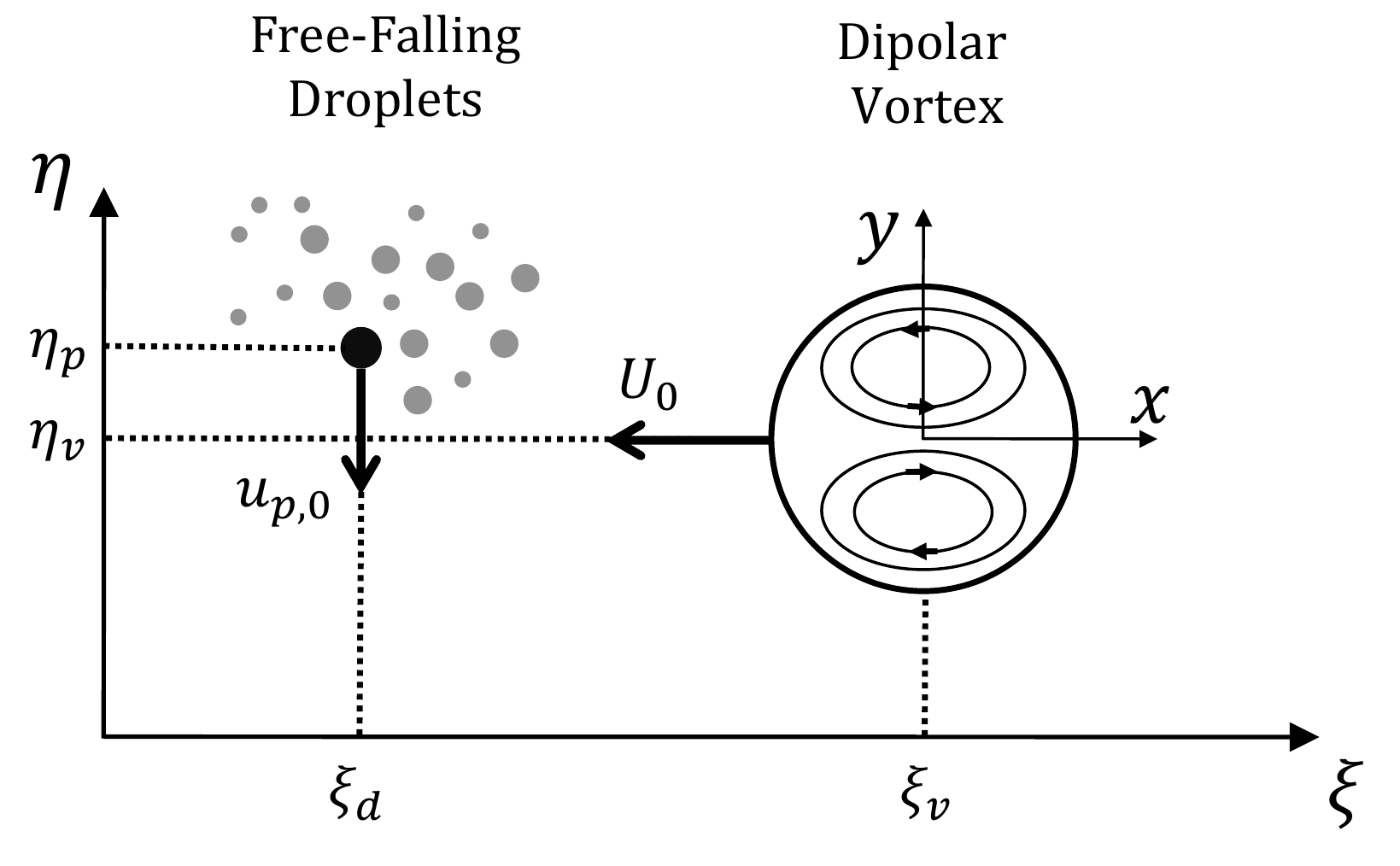}}
    \caption{Flow-particle configuration illustrated with respect to the stationary frame of reference $(\xi, \eta)$. Droplets, initially placed at $(\xi_p, \eta_p)$, free fall due to gravity $\bar{g}$. A vortex dipole flow-field initially placed at $(\xi_v, \eta_v)$, is self-propelling towards the free-falling droplets at a velocity $U_0$.}
    \label{fig:conf}
\end{figure}

A vortex dipole self-propelling in ambient air moving towards a cloud of free-falling droplets is illustrated in Fig. \ref{fig:conf}.
Initially, the vortex is located at $(\xi  _ {v}=0,\eta _ {v}=2m)$ in a stationary frame of reference, and translates with a velocity of $U_0$ parallel to the ground.
Naturally, the size of typical indoor wake structures is limited by the presence of walls, ceilings, and other boundaries, and thus we may investigate vortices having a diameter in the limit between $a=0.01m$ and $a=1m$.
Following the same logic, the propagation velocity was assumed to be roughly $U_0\approx 0.1-0.5 m/s$, representing a typical wake structure due to the motion of objects in a room, e.g., a hand movement \citep{Luo2018a}.

We analyze the influence of the vortex on the trajectory of a single Lagrangian droplet located $(\xi  _ {p}-\xi  _ {v},\eta _ {p}-\eta _ {v})$ relative to the vortex center, and is free-falling at a terminal velocity matching the droplet's initial diameter $d_{p,0}$,  
\begin{equation}
    \vec{u}_{p,0}=\frac{\rho_p d_{p_0}^2}{18\nu_f \rho_f}  \vec{g}.
\end{equation}
Since we aim to investigate the indoor dispersion of small saliva droplets, we consider the droplet's initial diameter in the range $d_{p,0}=1 - 200 \mu m $ \citep{smith2020aerosol}.
Moreover, we fix the environmental conditions and take the ambient air temperature as $T_{\infty}=298K$, the relative humidity as $RH=50\%$, and the droplet's initial temperature to be $T_{p,0}=308K$ for all the results presented in the following section.

% SECTION 3
\section{Results} \label{sec:results} 

\begin{figure*} 
    \centering
    \centerline{\includegraphics[width=\textwidth]{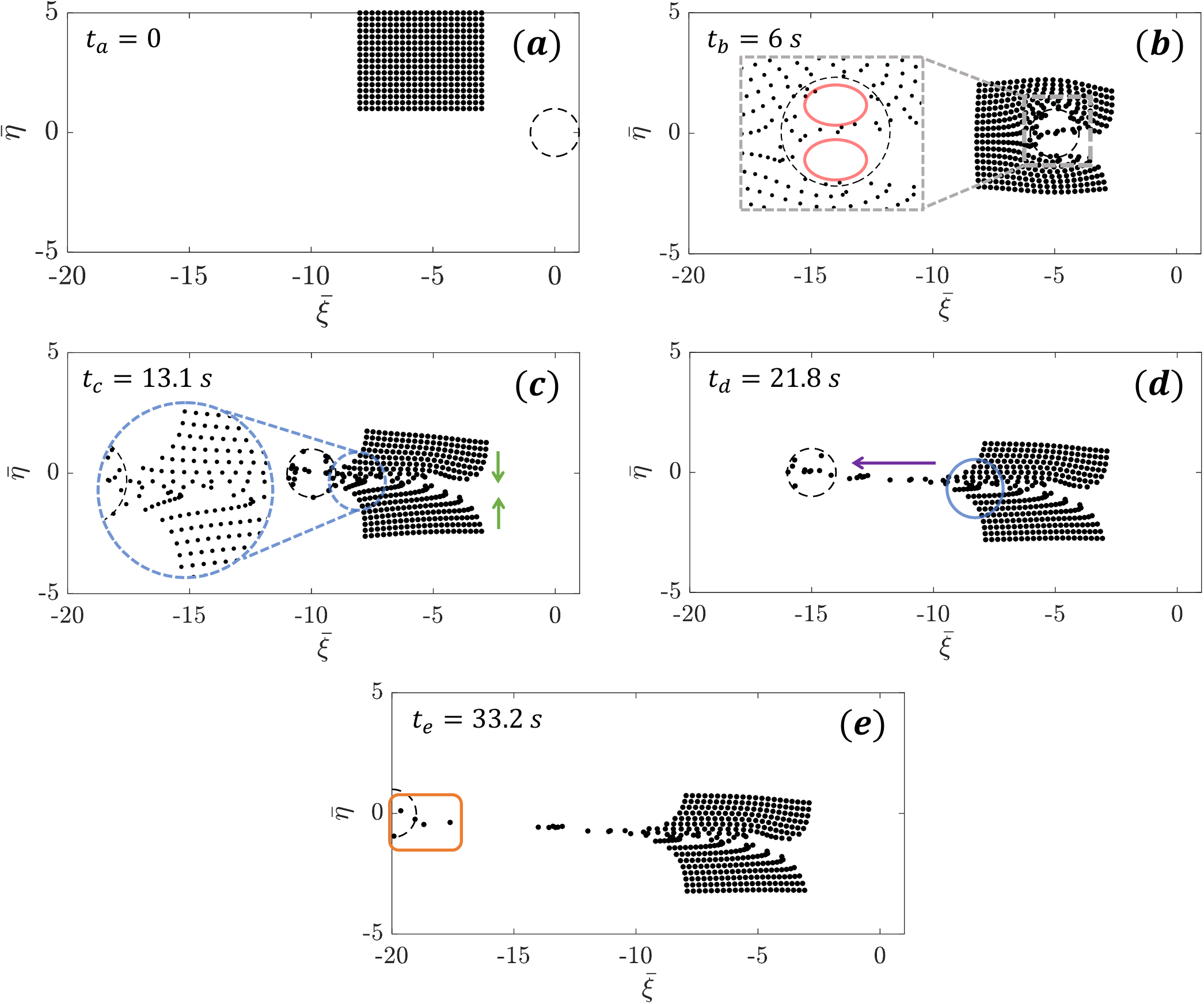}}
    \caption{ Instantaneous snapshots of the dispersion caused by a vortex of initial maximal intensity $\omega_{m,0}=10 s^-1$, radius of $a=0.1m$, and initial propagation velocity $U_0 \approx 0.1 m/s$ passing through a dilute cloud of $d_0=65 \mu m$ saliva droplets. The dashed line represents the vortex core, moving from right to left. Subplots (a-e) illustrate the cloud dispersion at different times. Both axes are normalized using the vortex radius $a$.  }
     \label{fig:cloud}
\end{figure*}

The dispersion of a droplet cloud by a passing vortex structure is illustrated in Fig.\ref{fig:cloud}, in a stationary frame of reference $(\bar{\xi},\bar{\eta})$; both axes are normalized by the vortex radius.
The vortex is chosen such that its initial maximal intensity is $\omega_{m_0}=10s^{-1}$, its radius is $a=0.1m$, and its initial propagation velocity $U_0 \approx 0.1 m/s$; such vortex may represent a characteristic wake flow structure.

To demonstrate the dispersion of droplets interacting with a passing vortex, the cloud is modeled as a two-dimensional array consisting of discrete $65\mu m$ saliva droplets, free-falling in the region $-8<\bar{\xi}<-3$, $1<\bar{\eta}<5$ outside the vortex core (initial state of Fig.\ref{fig:cloud}a). 
This array may serve as a general model for respiratory droplets falling in the ambient air after an exhalation event has occurred, as we may extend it further downstream without changing the essence of our results.

 Furthermore, the droplets are spread evenly as the distance between two neighboring droplets is one order of magnitude smaller than the radius of the vortex $0.1a$.
The spacing is much larger than the droplets' diameter, maintaining the dilution assumption; however, the macro-scale behavior of a cloud of droplets is still distinguishable, allowing us to understand the underlying mechanisms behind the cloud dispersion by the vortex. 

The evolution of the cloud formation is illustrated in Figures \ref{fig:cloud}b-e, presenting the scattering of the droplet cloud at different time instances.
Each time instance corresponds to the vortex translating $5a$ downstream.
Notably, since the dipole's transitional velocity decays, the time intervals chosen between each instance increase accordingly.

Fig.\ref{fig:cloud}b demonstrates the initial dispersion of the cloud.
When the vortex enters the cloud, it pushes most droplets in its immediate vicinity outwards and around it, creating dilute regions within the cloud (red circles), corresponding to the high-vorticity zones inside the dipole (Fig.\ref{fig:strm}).
The zero-vorticity centerline is the only region inside the dipole core remaining occupied, as the droplets arrange around these dilute zones.

The vortex keeps progressing and breaks through the cloud, as observed in Fig.\ref{fig:cloud}c. In contrast to the initial stage, when the vortex exits the cloud, the low pressure downstream pulls the droplets towards the vortex centerline.
This phenomenon is illustrated in the figure by the green arrows. 
While the general outer structure of the array does not alter substantially, the vortex entraps some droplets and carries them along when it moves away from the cloud. 
In turn, this dynamic alters the droplets' density inside the cloud and creates a distinct clustering region (blue circle, Fig.\ref{fig:cloud}c-d) near the vortex centerline. 
Although our model does not account for particle interaction, it is evident that droplets may cluster in certain regions.
There, droplets' collisions and interactions play an important role and may lead to the generation of larger, more virus-laden saliva droplets.

As the vortex propels downstream, a distinct droplet trace forms due to the gradual discharge of the entrapped droplets from the core, highlighted in Fig.\ref{fig:cloud}d (purple arrow). The trace maintains and outstretches even when the vortex passes far from the vicinity of the cloud, as illustrated in Fig.\ref{fig:cloud}e.
Ultimately, the complex nature of the cloud dispersion pattern is illustrated in Fig.\ref{fig:cloud}e; most of the droplets are nearly free-falling and settle vertically, whereas some droplets are advected by the vortex significantly, up to 20 times the size of the vortex core (orange frame in Fig.\ref{fig:cloud}e).
And, as will be shown later, the trapped droplets and aerosols may, in some situations, reach an extremely long settling distance.

\begin{figure}
    \centering
    \centerline{
    \includegraphics[width=\textwidth]{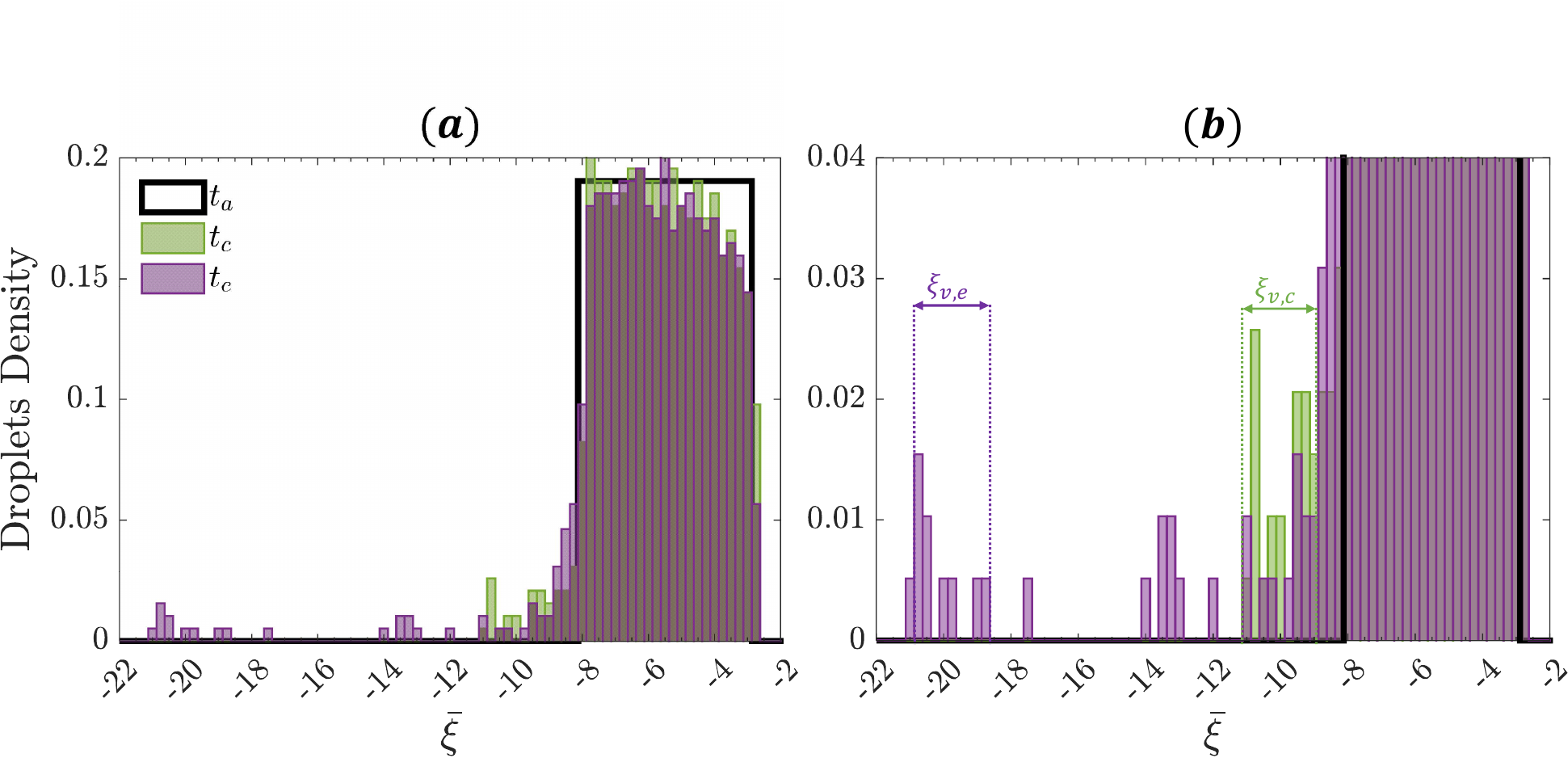}}
    \caption{Droplet spatial density as function of the normalized horizontal location at time instances $t_a=0s$ (black lines, uniform array), $t_c=13.1s$ (green bars, vortex temporal location $\bar{\xi}_v=-10$), and $t_e=33.2s$ (purple bars, vortex temporal location $\bar{\xi}_v=-20$), fitting the dispersion presented in Fig. \ref{fig:cloud}a,~\ref{fig:cloud}c, and \ref{fig:cloud}e, correspondingly. The vortex dispersion is illuminated by subplot (b), presenting the spatial density in a focused ordinate. The temporal location of the vortex is marked by dashed lines.}
     \label{fig:hist}
\end{figure}

 Following Fig.\ref{fig:cloud}, which uncovered the qualitative relation between the droplet's initial location and its dispersion, the dispersion of the cloud is quantified in Fig.\ref{fig:hist}.
The droplets' spatial distribution is depicted at three time instances; $t_a=0, t_c=13.1s, t_e=33.2s$, in which the center of the self-propelling vortex is located at $\bar{\xi}_{v,a}=0$,  $\bar{\xi}_{v,c}=-10$, and  $\bar{\xi}_{v,e}=-20$, correspondingly. 
The presented spatial distribution functions match the snapshots presented in Fig.\ref{fig:cloud}a, ~\ref{fig:cloud}c, and \ref{fig:cloud}e. 
As previously mentioned, initially ($t_a$) the droplets are evenly scattered in the region $-8<\bar{\xi}<-3$. 
Examining the density distribution at $t_c=13.1s$, the onset of the dispersion due to the vortex movement is evident, and we find that 11\% of the droplets displace to the edge of the initial cloud boundaries.
Some displaced droplets have translated along with the vortex core (centered at $\bar{\xi}_{v,c}=-10$, dashed green lines), while others remained in the vicinity of the cloud left boundary, and thus would not continue to translate inside the core.

The droplets' density distribution at the last time instance $t_c=33.2s$, where the core reaches the distance of $\bar{\xi}_{v,c}=-20$, reveals the full extent of the dispersion caused by the vortex passing the cloud; 22\% of the droplets have transported outside the initial boundaries. 
Most of the droplets' displacement was insignificant and led to a subtle downstream diffusion of the cloud. However, 5\% of droplets were entrapped inside the vortex core and carried along substantially, more than $20a$, as the vortex propelled leftwards.

The noteworthiness of the cloud dispersion unveiled in Fig.\ref{fig:hist} amplifies when considering the viral load transmitted by the droplets.
Coughing, sneezing, or speaking may produce large respiratory saliva droplets containing a sizeable number of virus copies but tend to settle quickly because of their relatively large mass. 
However, due to the characteristics of saliva, the evaporation of these droplets may generate fine aerosols, as presented in Fig.\ref{fig:val}.
Since the aerosols originated from much larger droplets losing their water content, and despite their small volume, they might carry an unanticipated large viral load.
On the other hand, the very same aerosols are the ones that are prone to entrapment inside the vortex core and might be propelled by these indoor wake flows.
Once trapped, these aerosols will remain inside the vortex for significant times due to their low inertia, up to the point where the viscid decay of the vortex diminishes, allowing their escape and subsequent settling. 
Thus, although Fig.\ref{fig:hist} suggests that merely 5\% of the droplets (turned aerosols) were trapped by the wake structure, they might boost the infection probability of nearby individuals, as they carry more virus copies compared to similar-sized aerosols generated during expiratory events.
For that reason, we seek to isolate and further investigate the variables dictating the extent of such dispersion.

\begin{figure}
    \centering
    \centerline{\includegraphics[width=0.55\textwidth]{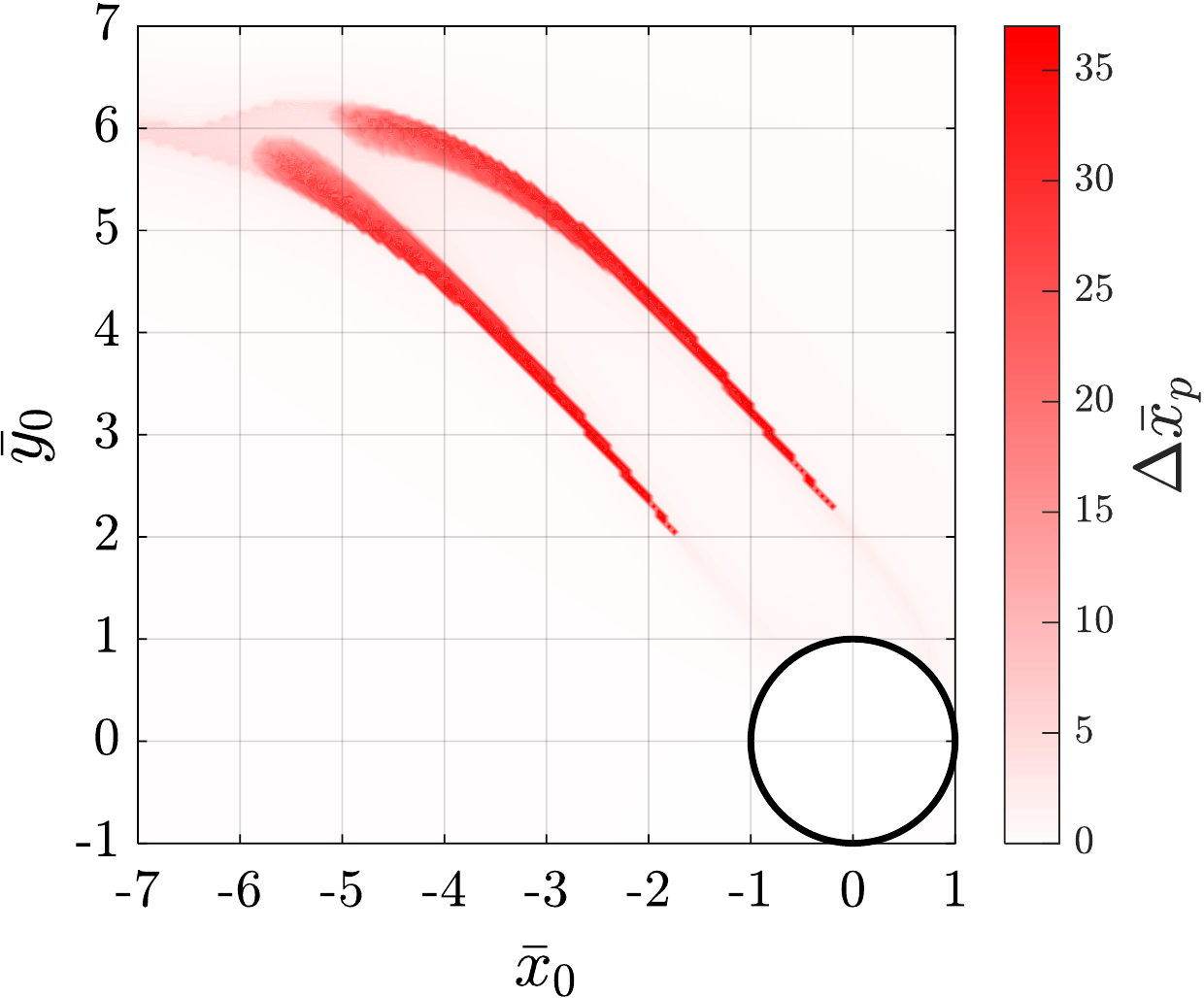}}
    \caption{Downstream settling distances of $d_0=65 \mu m$ saliva droplets dispersed by a vortex of $a=0.1m$ radius and $\omega _{m} =10s^{-1}$ initial maximal intensity, mapped as function of their initial location $(x_0,y_0)$ relative to the vortex location, denoted by a solid black line. The settling distance $\Delta x_p$, and both axes, are normalized using the vortex radius $a$.}
     \label{fig:map}
\end{figure}

The influence of the droplet's initial location $(x_0,y_0)$ relative to the vortex core on its downstream settling distance $\Delta x_p$, i.e. the difference between the droplet's initial horizontal location and its settling location, is quantified in the regime diagram plotted in Fig.\ref{fig:map}.
Here, the investigated flow configuration is similar to the one used in Fig.\ref{fig:cloud}, as well as the droplet's initial diameter.
Two distinct "critical traces" are visible, marking the initial locations of droplets that will eventually be carried to significant distances downstream, up to a maximum settling distance of $35a$.
Droplets located outside these regions, and notably between the two traces, are uninfluenced by the vortex, as they settle either ahead or after the vortex has passed their vicinity and thus free-fall almost undisturbed.

One shall observe that droplets located too close to the vortex, i.e. $y_0<2$, would not be translated significantly by the flow. 
This seemingly counterintuitive behavior originated from the droplet's evaporation process, as it dictates the penetration and escapes from the core. If the droplet reaches the vicinity of the core prematurely, it will not lose significant mass and have enough inertia to escape the core.
We note that the droplet evaporation timescale for the studied environmental conditions is $t\sim10s$ (see Fig.\ref{fig:wells} and Fig.\ref{fig:val}), matching its relaxation timescale $t={u}_{p,0}/a\sim10s$ and the vortex propagation timescale $t=U_{0}/a\sim10s$. 
Thus, critical regions appear above $y_0=2$, where the droplets that do penetrate the core are light enough to be transported considerably.
Curiously, from the inception of the traces and up to $y_0=5$, the transition between the regions is distinctly sharp, suggesting a slight perturbation in the droplet's location might result in a distinctly different droplet-vortex interaction. 
The traces seem to merge around $(-6,6)$ as the sharp transition fades, in correspondence with the decay of the vortex impact that quickly turns negligible for droplets located further downstream. 

\begin{figure*}
  \centerline{\includegraphics[width=\linewidth]{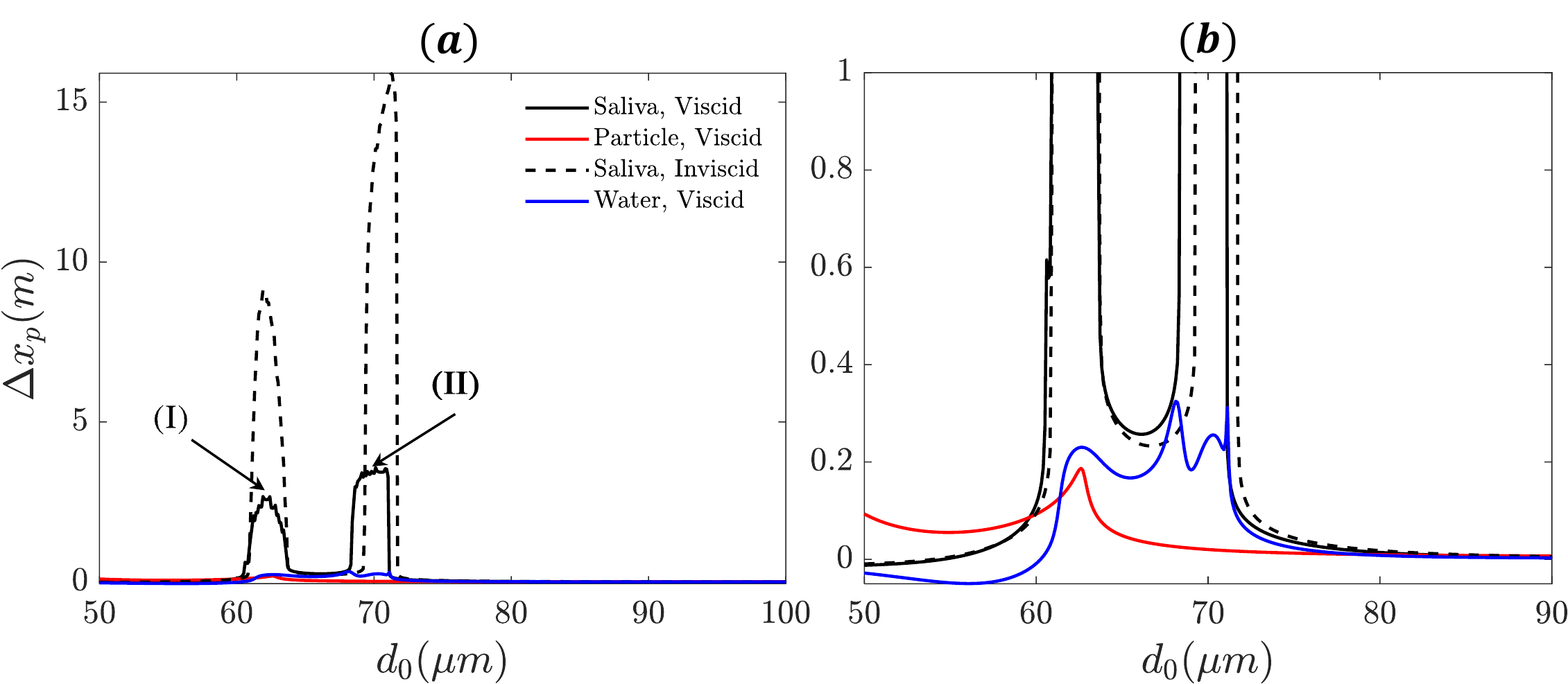}}
  \caption{ Downstream settling distances of saliva droplet (black line), water droplet (blue line), and solid particle (red line) with variable initial diameter, dispersed by a vortex of radius $a=0.1m$ and initial maximal intensity of $\omega _{m} =10s^{-1}$. The dashed line presents the dispersion generated by an ideal, inviscid vortex, whereas other solid lines mark the results of the viscid, decaying wake. All tracked particles were released from $(-3,3)$ relative to the vortex center. The double-critical behavior is illuminated by subplot (b), presenting the settling distances in a focused ordinate. }
  \label{fig:d0_delX}
\end{figure*}

The "double critical" result presented in Fig.\ref{fig:map} suggests that the dispersion of the saliva droplet by a wake structure is highly sensitive to its location relative to the disturbance origin, as slight changes may lead to vastly different settling distances. Considering that, We aim to analyze the dependence of the settling distance on the initial diameter of the droplets. We maintain constant initial location and wake flow characteristics. Fig.\ref{fig:d0_delX} explores the settling distances of various saliva droplets located at $(-3,3)$ and dispersed by the wake structure used previously.
We seek to estimate the extent to which the droplet's evaporation, the saliva salt composition, vortex decay, and their cross effects influence the droplet's downstream translation as well.
Thus, Fig.\ref{fig:d0_delX} also compares the settling distance of a saliva droplet with the settling of a water droplet (no salts), solid particle (no evaporation), and the settling of a saliva droplet dispersed by an ideal, inviscid dipolar vortex (no decay).

A prominent double-critical result is evident in Fig.\ref{fig:d0_delX} too.
Two sharp peaks of settling distances were observed for saliva droplets around $d_0=65\mu m$ and $d_0=70\mu m$, while droplets outside this narrow diameter range, either small or large, were not transported significantly by the dipole.
 This result suggests that the dispersion of the droplet by a wake structure is highly sensitive to the diameter of the dispersed droplet as well, owing to the complex entrapment mechanism.
Similar highly non-monotonic relation between the particle's size (or, equivalently, the Stokes number) and its dispersion was analyzed by Esmaily and Mani~\citep{Esmaily2020a,Esmaily2020b}. 
Notably, their analytic modeling revealed that the non-linearity exhibited by three-dimensional isotropic turbulence might be reproduced by an oscillating, one-dimensional flow.

The comparison between the solid particle, water droplet, and saliva droplet highlights the necessity of careful thermodynamic modeling since both the particle and the water droplet settling distances are of one order of magnitude smaller than the settling of the saliva.
In particular, this result might infer that accounting for saliva non-volatile contents is critical when analyzing the dynamics of exhaled respiratory droplets, as the aerosol generation process may significantly alter the dispersion pattern.
When considering the droplet-aerosol transition, the persistence of the aerosols results in much greater settling times, and now the effect of the wake viscid decay is conspicuous, as demonstrated in the dashed line in Fig.\ref{fig:d0_delX}. 
Albeit maximal translation is achieved for droplets of similar diameters in both viscid and inviscid cases, the viscid decay alters the aerosol ability to remain entrapped inside the core. Thus, the settling distance predicted for the ideal case might be an overestimation.
Unsurprisingly, When the droplet-vortex interaction is not dominant, both wake models yield a similar settling distance prediction, further highlighting the strong coupling between the settling distance and the aerosol's residence time inside the vortex.

\begin{figure*}
\centerline{\includegraphics[width=\linewidth]{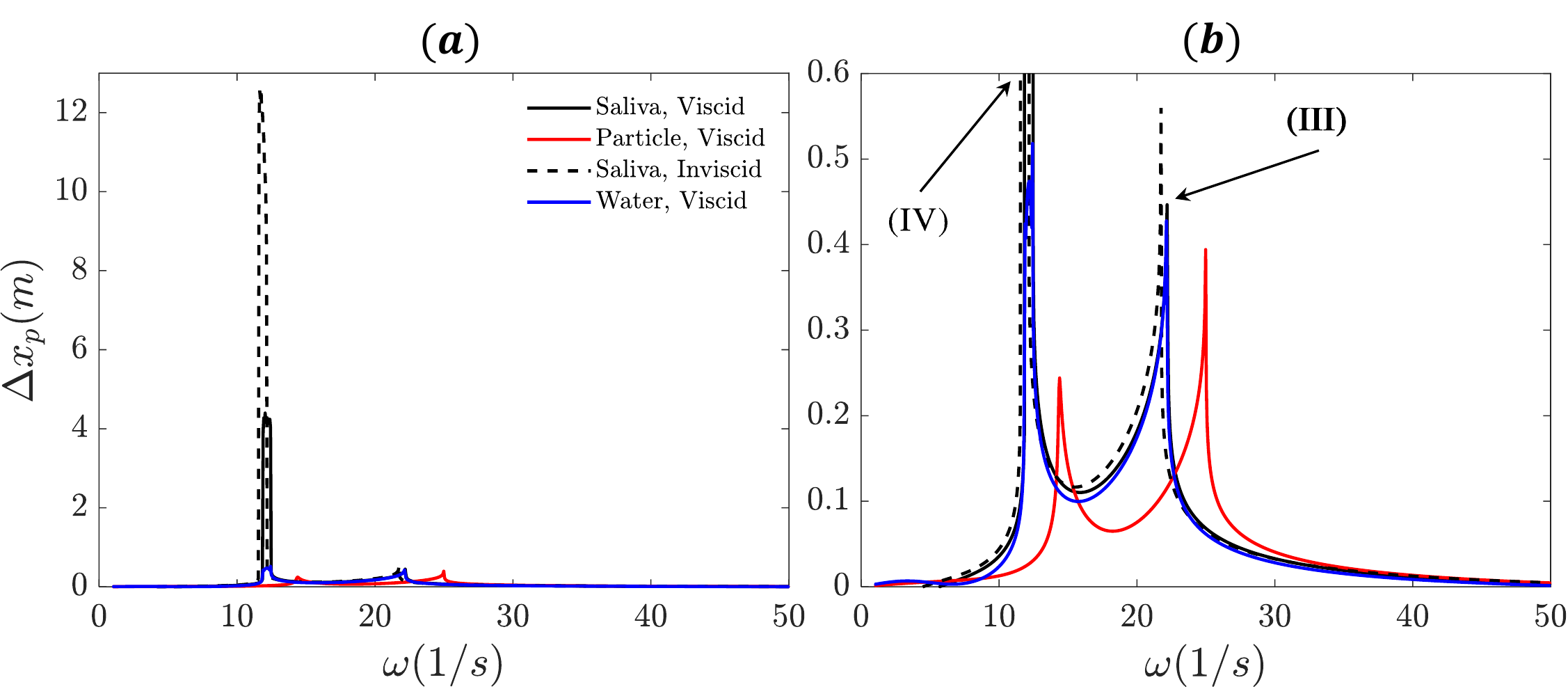}}
    \caption{ Downstream settling distances of $d_0=75 \mu m$ saliva droplet (black line), water droplet (blue line), and solid particle (red line), dispersed by a vortex of $a=0.1m$ radius and variable initial maximal intensity. The dashed line presents the dispersion generated by an ideal, inviscid vortex, whereas other solid lines mark the results of the viscid, decaying wake. All tracked particles are released from $(-3,3)$ relative to the vortex center. The double-critical behavior is illuminated by subplot (b), presenting the settling distances in a focused ordinate. }
    \label{fig:w_delX}
\end{figure*}
\begin{figure*}
\centerline{\includegraphics[width=\linewidth]{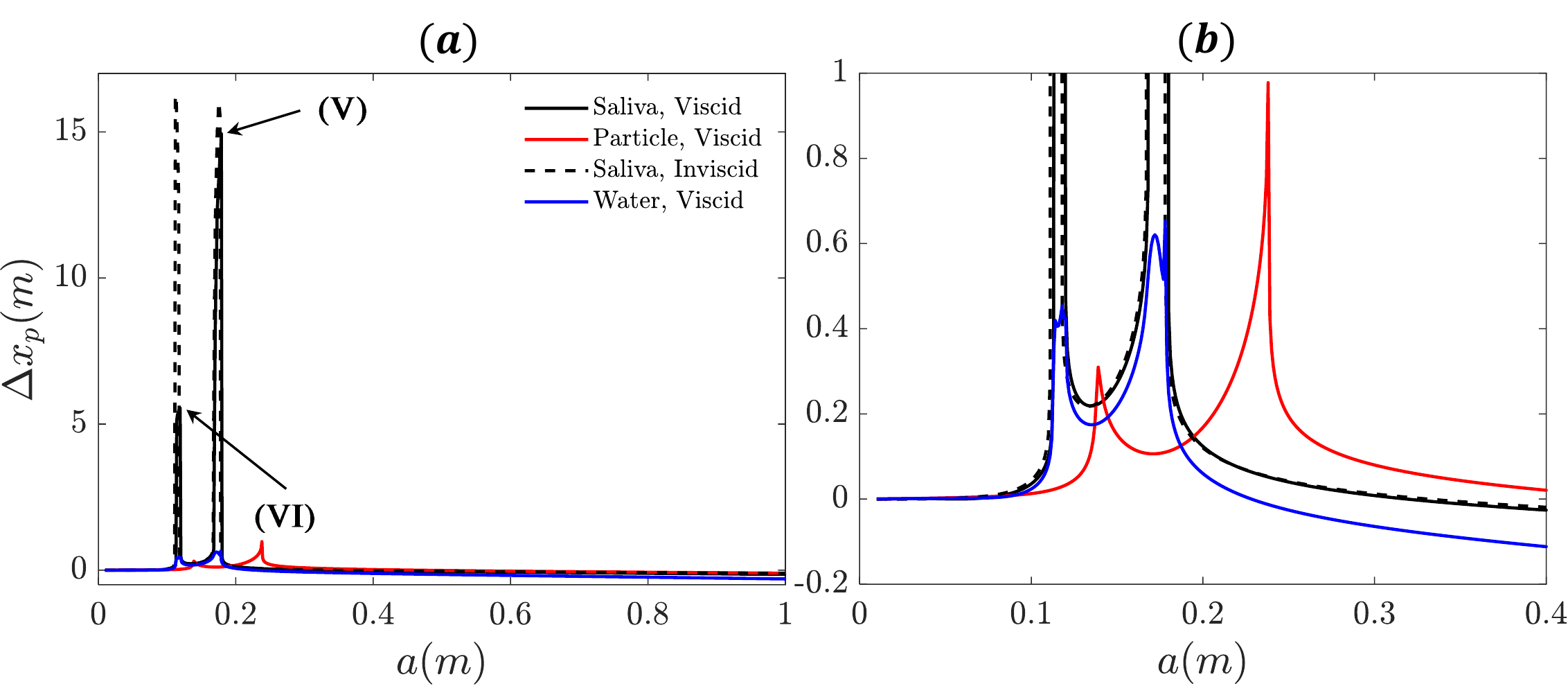}}
    \caption{ Downstream settling distances of $d_0=75 \mu m$ saliva droplet (black line), water droplet (blue line), and solid particle (red line), dispersed by a vortex of $\omega _{m} =10s^{-1}$ initial maximal intensity and variable radius. The dashed line presents the dispersion generated by an ideal, inviscid vortex, whereas other solid lines mark the results of the viscid, decaying wake. All tracked particles were released from $(-3,3)$ relative to the vortex center. The double-critical behavior is illuminated by subplot (b), presenting the settling distances in a focused ordinate. }
    \label{fig:a_delX}
\end{figure*}

Finally, the wake flow influence is analyzed by studying the settling of particles located at $(-3,3)$ with an initial diameter of $d_0=75 \mu m$. 
Two cases are studied; in Fig.\ref{fig:w_delX}, the vortex radius was kept constant $a=0.1m$ as the values of the vortex initial intensity $\omega _{m,0}$ were changed, whereas in Fig.\ref{fig:a_delX}, the vorticity was set as $\omega _{m,0}=10s^{-1}$ and different vortex radii were studied. 
For both cases, and in correspondence to previous results, the saliva droplets were translated substantially farther than the water droplets or the solid, non-evaporating particles. Moreover, the expected double critical relations were obtained for both the droplets and the solid particles, but unlike Fig.\ref{fig:d0_delX} the two peaks observed in Fig.\ref{fig:w_delX} and Fig.\ref{fig:a_delX} are distinctly uneven.
One may observe that the introduction of non-volatile components does not change the values (vorticity or radius) for which maximal displacement is achieved, evident in the overlap of the peaks for saliva and water droplets.
On the other hand, we find that non-volatile particles may reach the maximal settling distances when dispersed by vortices of larger radii or higher maximal vorticity, resulting in a rightward shift of the curve.

Fig.\ref{fig:w_delX} reveals that low-intensity vortices do not affect the droplets significantly, since the flow fields they induce are too weak to overcome the droplet inertia. 
However, at the limit of high intensity, vortices did not scatter the droplet farthest, as perhaps expected.
Although such dipoles might be more energetic, their translation velocities $U$ are, accordingly, higher.
As a result, the strong vortices pass the vicinity of the droplet rapidly, not allowing for the particles and droplet to entrap inside the core.
Thus, overall, increasing the vorticity might decrease the settling distance of the droplets. 
Notably, critical translation is attained only at a narrow strip of initial vorticities for a given configuration, suggesting that unstable, critical particle-vortex interaction is linked to one specific vortex frequency.
The intricacy of the droplet-aerosol transition is emphasized here; the strong linking between the displacement and a single vortex frequency is maintained only for evaporating saliva droplets, indicating the drying process might be the origin of the imbalance apparent in Fig.\ref{fig:w_delX}.

Similarly, Fig.\ref{fig:a_delX} unveils another non-trivial dynamic and suggests that large enough vortices translate the droplets contrary to their propulsion direction.
Larger flow structures are more powerful and might disperse the droplets farther; nevertheless, vortices larger than $a=0.25m$ seem to drive the droplets opposite to its trajectory, as the flow outside the dipole carries the droplet away from the core before it reaches its vicinity, thus eliminating the possibility of entrapment and subsequent downstream translation.

The analytic relation between the size of the dipole core and its viscid decay rate is given by Eq.\ref{eq:dec}, suggesting that smaller vortices may decay faster. 
Fig.\ref{fig:a_delX} reflects the ramification of such decay, which is the root cause of the unevenness in the critical conditions.
For the ideal flow two, almost equal peaks were found, for radii $a=0.12$ and $a=0.18$, as the saliva droplets were displaced roughly 15 m downstream.
When considering the dissipation of the dipole, albeit most of the results are unchanged, the settling distance at the left peak, corresponding to the smaller diameter $a=0.12$, was reduced to a third of the dispersion for an inviscid wake. 
Thus, one may conclude that a careful examination of both flow and particle deposition characteristic times is necessary when estimating the validity of particle-wake flow interactions.
The effect of the viscid dissipation shall be negligible when considering larger vortical structures but may alter the particle dynamics significantly when even slightly smaller structures are investigated.

\begin{figure*}
    \centering
    \centerline{\includegraphics[width=\textwidth]{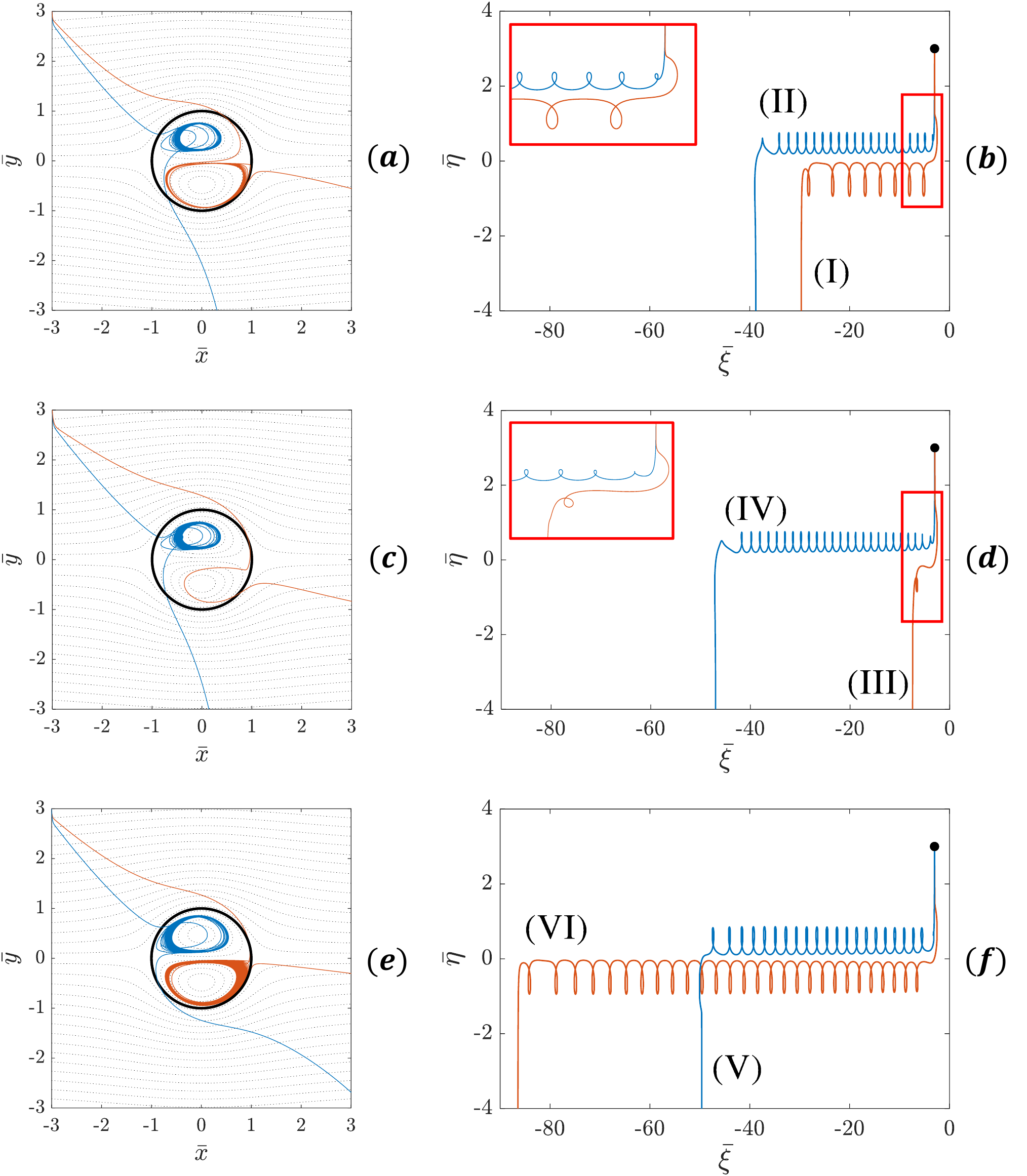}}
    \caption{Comparison of evaporating saliva droplets' critical trajectories. Left-hand side subplots (a,c,e) present the trajectories of the droplets in the vortex frame of reference, while the right-hand side subplots (c,d,f) show the same trajectories relative to the ground. The conditions for which each trajectory is found are listed in Table \ref{tab}.}
    \label{fig:critical}
\end{figure*}

The droplet's critical trajectories (labeled I-VI in Fig.\ref{fig:d0_delX}-\ref{fig:a_delX} and Table \ref{tab}) are presented in Fig.\ref{fig:critical}, in both stationary (left subplots) and a frame of reference following the vortex center (right subplots).
\begin{table}[h]
\centering
    \begin{tabular}{c@{\hspace{0.5in}}c@{\hspace{0.2in}}c@{\hspace{0.2in}}c}
    Trajectory & $d_0~(\mu m)$ &$\omega_{m,0}~(1/s)$& $a~(m)$ \\ \hline
    I & 62 & 10 & 0.1 \\   
    II & 70 & 10 & 0.1 \\ 
     III & 75 & 12 & 0.1 \\ 
      IV & 75 & 22.7 & 0.1 \\ 
       V & 75 & 10 & 0.12 \\ 
        VI & 75 & 10 & 0.18 \\ \hline
    \end{tabular}
    \caption{Conditions for which critical trajectories are attained for saliva droplets, as reported in Fig.\ref{fig:d0_delX}-\ref{fig:a_delX}.}
    \label{tab}
\end{table}

All the trajectories illuminate the droplets' substantial transport, which is a result of droplet entrapment within the vortex core.
Droplets that had sufficient inertia to enter the core lost mass due to evaporation while rotating inside the vortex core and, as a result, could not escape. 
In this case, the droplet-aerosol transition may occur when the particle is within the vortex core, and thus it is carried along with the vortex downstream, leading to substantial and unanticipated translation.
This motion is demonstrated in the right subplots of Fig.\ref{fig:critical}, where most of the critical trajectories exhibit distinct circular motion, tracing the motion of the self-propelling, circulating dipole. 
Curiously, although significant differences exist between the trajectories, they all penetrate the vortex core at the same locations; either near the right (I, III, and V) or the left  (II, IV, and VI) stagnation point. This result might clarify the origin of the double critical relations aforementioned.
The passing of the droplet near the stagnation point allows it to penetrate the core and circulate inside it, due to lower relative velocity in the stagnation region.
We observe two critical trajectories, fitting each stagnation point.
Remarkably, any slight changes in the droplet's initial location and diameter, or in the vortex intensity and size, might divert the droplet away from the stagnation region, hence significantly reducing its translation by the vortex.

To highlight this realization, we shall investigate trajectory I.
There, the droplet penetrates the core at the back stagnation point and hence we may conclude that a lighter droplet would have washed downstream and thus influenced by the vortex only briefly.
Following the same logic, a heavier droplet would have entered the core earlier and closer to the vortex maximal vorticity point.
Consequently, it would not have crossed the vicinity of the stagnation point, but rather swirl inside the core and eject rapidly, leading to a sharp decrease in the settling distance.
Further increase in the droplet mass will move the penetration point closer to the front stagnation point (trajectory II) and once again the translation of the droplet is magnified, as it is trapped and swirls around the rotating pole.
We may apply the same analysis for the rest of the results, therefore offering a general interpretation of the findings presented in Fig.\ref{fig:d0_delX}-\ref{fig:a_delX}.

The abovementioned rapid ejection is evident in trajectory III, thus explaining the dominant disparity observed in Fig.\ref{fig:w_delX}.
Due to higher vorticity, every droplet that penetrates the core does not lose a substantial amount of mass as it swirls around the core quickly, thus having enough inertia to cross the carrier flow streamlines and escape the core.
Henceforth, only low-vorticity wake flow manages to transport the saliva droplets significantly, as the characteristic residence time inside their core is higher, which allows entrapment.

Finally, a careful examination of the critical trajectories divulges the effect of viscid decay.
For example, the decline in the intensity of the dipole $\omega_{m,0}$ manifests in trajectory V.
When considering the trajectory in a stationary frame of reference, we observe that the spacing between two neighboring loops grows continuously, up to the point where aerosols escape.
This indicates that the particle's swirl frequency decays, as expected; the characteristic frequency of the vortex is directly proportional to the particle's swirl frequency.

%SECTION4
\section{\label{sec:conc} Conclusions}

A mathematical analysis of the dynamics and dispersion of exhaled saliva droplets under the influence of an approaching dipolar vortex was carried out.
Our analysis reveals complex dynamic interactions of the evaporating droplets and the induced wake flow,  significantly affecting the settling distances of the micron-sized expiratory droplets.
Once captured inside the vortex core, the droplet might rotate within the vortex core and translate downstream along with the self-propelling vortex.
Trapped droplets exhibited exceptionally large settling distances, up to 90 times the vortex core length scale.
Droplets that did not penetrate the core or have quickly ejected out of it settled much closer to their initial locations.
The effect of evaporation and droplet composition was analyzed and studied by comparing the dispersion and settling distances of an evaporating saliva droplet, a water droplet, and a solid particle.
Insights on the influence of the wake flow were gained by comparing the dispersion caused by an ideal, inviscid carrier flow to a more realistic decaying wake.
Furthermore, different sizes and vortex intensities were considered, analyzing the resultant variance in the settling distances.

We found that the settling distances of saliva droplets dispersed by a vortical wake flow are highly sensitive to the flow and particle properties due to the highly non-linear coupling between the flow and the droplets. Moreover, the non-linearity is exacerbated when accounting for the non-volatile saliva components, giving rise to aerosol formation mechanisms.
Our study discovered that the aerosol generation process might alter the dispersion patterns to such an extent that accounting for the non-volatile contents may lead to fundamentally different dispersion and settling behavior.
The importance of the phenomenon is magnified when considering that the aerosols might carry an unanticipated large viral load as they originated from much larger droplets, which initially contained a sizeable number of virus copies.

The proposed model might offer a new analytic tool for assessing the spreading mechanisms of airborne-carried pathogens, such as SARS-CoV-2.
Taking this approach has allowed for the isolation of the interaction between a droplet and the vortical flow, thus aiding in uncovering the basic principles behind the reported enhanced transport of droplets by various vortical flows.
On the other hand, the current model does not account for turbulence effects, which may play a role in the dispersion and clustering of particles.
Hence, the extension of our approach in incorporating a random-walker model into the analytically-described carrier flow is currently being examined; we aim to study the influence of turbulent dispersion on the complex droplet dynamics revealed in the current study.
Moreover, broader heat and mass convection models could be incorporated into our model, while the changeable environmental conditions that might influence the dispersion mechanism, could be investigated as well. 

\section*{Declarations}

\bmhead{Acknowledgments}

This research was supported by the ISRAEL SCIENCE FOUNDATION (grant No. 1762/20).

\bmhead{Conflict of interest}

The authors declare that they have no conflict of interest.

\bmhead{Availability of data and materials}

All code used to obtain the numerical results of the present study is available from the authors upon reasonable request.

\bibliography{Final.bib}
\end{document}